\documentstyle[aps,prl,multicol,epsf]{revtex}

\begin{document}
\draft
\title{Ground State Spin Structure of Strongly Interacting Disordered 
1D Hubbard Model}

\author
{E. Eisenberg and R. Berkovits}

\address{
The Minerva Center for the Physics of Mesoscopics, Fractals and Neural 
Networks,\\ Department of Physics, Bar-Ilan University,
Ramat-Gan 52900, Israel}

\date{\today}
\maketitle

\begin{abstract}
We study the influence of on-site disorder on the magnetic properties
of the ground state of the infinite $U$ 1D Hubbard model. We find that
the ground state is not ferromagnetic. This is analyzed in terms of the 
algebraic structure of the spin dependence of the Hamiltonian. A simple 
explanation is derived for the $1/N$ periodicity in the persistent current 
for this model.
\end{abstract}

\begin{multicols}{2}
\narrowtext

The Hubbard model is the simplest model one can study to examine the 
effects of correlations between electrons in narrow energy bands.
The Hamiltonian consists of a nearest neighbor hopping term and an 
electron-electron repulsion $U$ which acts only when two electrons are on 
the same site. The Hubbard model is also the canonical model for the 
study of itinerant ferromagnetism. The strong coupling regime is of 
special importance for the study of ferromagnetism, since a theorem by 
Nagaoka \cite{nagaoka} states that in the $U=\infty$ limit, the GS
is ferromagnetic given some connectivity property of the lattice 
(which holds in most cases for $d>1$). The model is solvable in 1D and it
was shown \cite{lieb} that for open boundary conditions (BC) the ground
state (GS) for finite $U$ is a singlet, i.e., there can be no ferromagnetism 
unless one postulates explicitly spin- or velocity-dependent forces.
For infinite $U$, the GS of all spin sectors are degenerate.

The problem of the interplay between disorder and interactions in 
systems of electrons is challenging and has a long history \cite{review}.
It is of interest to study the influence of disorder on the possibility 
of forming a ferromagnetic GS. In this work we study the spin structure 
of the GS of a disordered $U=\infty$ Hubbard model in 1D. We find that 
for periodic BC as well as for open BC, for any realization of on-site
disorder, the GS is degenerate, where all spin sectors have the same lowest 
energy, except for the fully polarized one which has a higher energy. 
As a by-product of our proof, we
find that the GS of an even (odd) number of spinless fermions on a 1D
ring threaded by flux, is minimal when the dimensionless flux $\Phi/\Phi_0$ 
equals $\pi$ ($0$). This might be of interest for the study of persistent
currents in disordered interacting 1D rings \cite{dipc}.

Lieb and Mattis\cite{lieb} have considered the one dimensional clean
Hubbard model for any $U<\infty$ given hard wall (or open) BC.
They found that the GS is always a singlet (for an even 
number of spins). When $U=\infty$ the GS in all the different spin 
sectors become degenerate. Here we present an analysis of the $U=\infty$
with periodic BC, in the presence of on-site disorder. 
As will become clear later, the different BC
change the character of the problem, and the periodic BC case give
us some insight into the higher dimensional variants of the problem.
The Hamiltonian we consider is thus given by 
\begin{equation}
H=\sum_{i\sigma} \varepsilon_i n_{i\sigma}
-t\sum_{i\sigma}a_{i\sigma}^\dagger a_{(i+1)\sigma}
+  c.c. +U\sum_i n_{i\uparrow}n_{i\downarrow},
\label{hamil}
\end{equation}
where $a_{i\sigma}^\dagger$ is the fermionic creation operator on site
$i$ with spin $\sigma$, $n_{i\uparrow}=a_{i\uparrow}^\dagger
a_{i\uparrow}$, and the on-site energies $\varepsilon_i$ are drawn
uniformly between $-W/2$ and $W/2$.

The Hilbert space is composed of a direct product of the spatial 
wave functions, described by a basis composed of all the different 
possibilities of positioning $N$ particles on $N$ out of the
$M$ different sites. This space is isomorphic to
the Hilbert space of $N$ non interacting spinless fermions on $M$ sites.
For finite $U$, the Hilbert space consists of functions with double
occupancy as well, and these exists only if the two particles on
the same site are in a singlet configurations. Thus, coupling between
the spatial functions and spin functions is formed. However, in the
$U=\infty$ case the spin Hilbert space is decomposed from the spatial
Hilbert space, and its natural basis is given by the $2^N$ orderings
of the $N$ spins.

The action of the hopping terms in the Hamiltonian on a wavefunction
is changing the spatial distribution of the particles, through moving
one particle at a time to one of its neighboring sites. Since double
occupancy is restricted, particles can not interchange their order
via hopping. One gets another invariant of the Hamiltonian, namely,
the order of the particles, the importance of which will become clear
shortly. This holds in the hard wall BC case.
However, the periodic BC allow particles to
''bypass'' through the boundaries and breaks this symmetry.
Had the ordering of the spins been conserved, the GS would have been
independent on the entire spin structure of the states. Consequently, 
in the hard wall BC case, all spin configurations GS are degenerate.
However, the change in the spin ordering
due to hopping through the boundaries, couples the spin part to the
spatial part of the wavefunction. Each hopping term in the spatial part
of the Hamiltonian is replaced by a matrix (in principle, $2^N\times 2^N$)
which characterizes the re-ordering of the spins. Most of these matrices,
namely, these which are not related to the boundary terms,
are identity matrices. However, some describe the re-ordering due to
hopping from site $1$ to site $N$, and vice versa, and are not trivial.

Let $T$ be the permutation in the spin ordering due to hopping of a
particle at site $1$, to site $N$. The minimal subgroup of the permutations
group which contains $T$ (and $T^{-1}$, which corresponds to hopping
in the inverse direction) is the cyclic group generated by $T$.
The different spin orderings induce a representation of this cyclic group.
Since the cyclic group is commutative, its irreducible representations
are one dimensional. Thus we will find the decomposition of the induced
representation into irreducible parts, and then diagonalize the
spatial Hamiltonian for each of the irreducible representations. The GS
is the lowest of the eigenvalues obtained for each representation.
Our goal is to determine to which $S$ value corresponds the 
irreducible representation with the lowest eigenvalue.

We note that since $T^N=I$, all the representations must be of the form
$\chi_j(T^k)=\exp(2\pi i jk/N)$. Thus, different representations correspond
to assigning a phase $\chi_j(T)=2\pi j/N$ to the transition through 
the boundaries. This is equivalent to the adding a flux $\Phi=2\pi j/N$ 
through the ring to the $N$ spinless fermions problem. 
We now prove the following

\bigskip
\noindent
{\underline{\it Theorem}}:
Let $\epsilon(\Phi)$ be the GS of $N$ non interacting spinless particles 
on a (disordered) ring, as a function of the flux $\Phi$.
$\epsilon(\Phi)$ has a minimum at $\Phi=\pi$
for even number of particles (and at $\Phi=0$ for odd $N$).

\noindent {\it Proof}:
The energy as a function of the flux $\epsilon(\Phi)$, is symmetric
in the parameter $\Phi$ with respect to the points $\Phi=0,\pi$.
Therefore, the first derivative $\epsilon'(\Phi)$ vanishes
in these points. We now calculate  the second derivative at these
points in order to find which is the maximum and which is the minimum.
Expansion of $\epsilon(\Phi)$ in the vicinity of $\Phi=\pi$ is given by
perturbation theory. The flux dependence of the matrix elements is
always through the expression $e^{i\Phi}$, and therefore one obtains
\begin{equation}
\epsilon(\Phi)-\epsilon(\pi)=\sum_n a_n (1-e^{i(\Phi-\pi)})^n.
\end{equation}
The curvature, i.e., second derivative at $\Phi=\pi$ is thus given by
\begin{equation}
\frac{\partial ^2\epsilon}
{\partial\Phi ^2}{\Big |}_{\Phi=\pi} = a_1-2a_2.
\end{equation}

At the point $\Phi=\pi$, all the off diagonal elements in the
Hamiltonian are negative, since the $-1$ factor of the flux which
multiplies the boundary hopping matrix elements cancels with the
$(-1)^{N-1}$ factor from the interchange of the fermionic operators.
Thus the GS eigenvector $v_0$ has no nodes, i.e., all its elements
are of the same sign. Now, let us write explicitly $a_1$ and $a_2$
\begin{equation}
a_1=\langle v_0| \Delta H |v_0\rangle = \sum_{[ij]} g_i g_j > 0,
\end{equation}
\begin{equation}
a_2=\sum \frac{|\langle v_n|\Delta H|v_0\rangle|^2}{E_0-E_n}<0.
\end{equation}
The sum over $[ij]$ is restricted to states $i$ and $j$ which are
coupled by boundary hopping terms.

It then follows that the curvature at $\Phi=\pi$ is positive, 
and therefore it is a local minimum point. 
Unless there is an additional accidental point
of vanishing derivative, this is also the global minimum.
Similar arguments apply to the point $\Phi=0$ for odd $N$. Q.E.D.

Note that the above theorem holds for off-diagonal disorder as well,
as long as all the hopping integrals are of the same sign.

One concludes that for even $N$, the irreducible representation which
gives the GS is $\chi_{N/2}$, and $\chi_{N/2}(T)=\chi_{N/2}(T^{-1})=-1$.
We now have to find out for which $S$ values the induced representation
includes $\chi_{N/2}$ in its decomposition. 
There is only one state with $S=S_z=S_{\rm max}$, and thus the induced 
representation is one dimensional and irreducible. It is easy to see that 
it is $\chi_0$ (which corresponds to the maximum energy). 
It will be now shown that the representation
induced by $S=0$ always include $\chi_{N/2}$.

Since the explicit construction of states with definite $S$ and $S_z$
is non trivial, let us look at the representations induced by the sets
of states with definite $S_z$ only. The irreducible representations
induced by $S_z=M$, which are not induced by $S_z=M+1$ correspond
to $S=M$. According to character theory, the number of times an irreducible
representation with character $\chi$ exists in the decomposition
of a representation $\phi$ is given by ${1\over K}\sum_k \phi(k)\chi^*(k)$,
where the index $k$ runs through all the group elements, and $K$ is the
group order. The character of a representation is just the trace of the
representing matrices. The character of the representation
induced by a set of states
is given by the sum over $k$ of the number of states in the relevant
set which are invariant under the $k$th element of the group. 
Accordingly, one obtains that the number of occurrences of $\chi_{N/2}$
in the representation induced by $S=1$ is given by 
\begin{equation}
k_1=\left\{
\begin{array}{lr}
\frac{(N-1)!}{(N/2+1)!(N/2-1)!}&  N/2\ {\rm even} \\ 
\frac{(N-1)!}{(N/2+1)!(N/2-1)!}-\frac{(N/2)!}{N((N+2)/2)!((N-2)/2)!} & 
N/2\ {\rm odd}
\end{array}\right\delimiter 0 ,
\end{equation}
and the number of occurrences
in the representation induced by $S=0$ is bounded by
\begin{equation}
k_0 \leq \frac{(N-1)!}{(N/2)!(N/2)!}. 
\end{equation}
It then follows that
$k_0 > k_1$, i.e., there exists at least one occurrence of
$\chi_{N/2}$ induced by $S=0$ states. Therefore, the GS is obtained
in the $S=0$ sector.

We have shown that for each (even) $N$, the GS is obtained at the $S=0$
sector, while the lowest energy in the $S_{max}$ sector is higher.

The above solution is based on the fact that the permutations induced
by the one dimensional boundary hopping terms spans only the cyclic group,
which is commutative. The representation induced by
the spin states is highly reducible with respect to this subgroup.
In higher dimensions, reordering of spins is
generated not only by boundary terms, and the whole (non commutative)
permutation group is needed to characterize the spin dependence.
It can be shown that the representation induced is irreducible with
respect to the full non-commutative group. Thus, the spin dependence
of the Hamiltonian is not equivalent to a trivial flux-like correction.

It is now very easy to consider the effect of magnetic flux added to the 
Hamiltonian. As we have seen, for an even number of electrons, the optimal 
total flux (physical flux + flux added by spin configuration) is (in 
dimensionless units) $\pi$. This indeed is the fictitious flux generated by 
the spins in the GS of the Hamiltonian without an external flux. Now, when we
turn on the magnetic flux, the GS energy will increase. It is therefore
favorable for the system to produce an inverse fictitious flux to cancel
out the influence of the magnetic flux. However, as we have seen, this
fictitious flux comes in quanta of $2\pi\Phi_0/N$, where $N$ is the number
of particles. Thus magnetic flux of integer multiples of $2\pi\Phi_0/N$ 
can be completely canceled out by the spin ordering, such that the GS energy 
is exactly as it were in the absence of flux. Therefore, as one increases 
the magnetic flux, the GS energy raises, up to the point $\Phi=\pi\Phi_0/N$,
where it is favorable for the system to produce a negative fictitious flux 
such that the total flux is (in absolute value) less than the magnetic flux.
We thus have a simple and transparent explanation for the flux dependence 
of the GS energy which is periodic with period $2\pi/N$ instead of the usual 
$2\pi$ period \cite{1dflux}.

In conclusion, we have shown in this work that the GS of spin $1/2$ 
fermions on a 1D ring is not polarized in the $U=\infty$ limit, for 
any realization of disorder. This is accounted for by mapping the 
spin background of the 1D problem onto a fictitious flux. In 2D the 
influence of the spin background is non-commutative and therefore much 
more complex. This leads to ferromagnetic GS for one hole \cite{nagaoka} 
and to disorder induced ferromagnetism for higher hole concentration 
\cite{prev}. It was also shown that the energy of spinless fermions is 
minimized when the applied magnetic flux is half the flux period, again, 
for any realization of disorder. This fact is relevant for persistent 
currents calculations.

\end{multicols}
\end{document}